\documentstyle[11pt]{article}
\begin{document} 

\makeatletter
\@addtoreset{equation}{section}
\makeatother
\renewcommand{\theequation}{\thesection.\arabic{equation}}
\baselineskip 15pt

\newtheorem{defofentangidentical}{Definition}[section]
\newtheorem{factorizabilityidentical}{Theorem}[section]

\title{\bf Teleportation with indistinguishable particles.\footnote{Work
 supported in part by Istituto Nazionale di Fisica Nucleare, Sezione di
 Trieste, Italy}}
\author{Luca Marinatto\footnote{e-mail: marinatto@ts.infn.it}\\
{\small Department of Theoretical Physics of the University of Trieste, and}\\
{\small Istituto Nazionale di Fisica Nucleare, Sezione di Trieste, Italy,}\\
and \\
\\ Tullio Weber\footnote{e-mail: weber@ts.infn.it}\\
{\small Department of Theoretical Physics of the University of Trieste, and}\\
{\small Istituto Nazionale di Fisica Nucleare, Sezione di Trieste, Italy.}}

\date{}

\maketitle

\begin{abstract}
We analyze in a critical way the mathematical treatment of a quantum
 teleportation experiment performed with photon particles, showing that a
 symmetrization operation over both the polarization and spatial degrees of
 freedom of all the particles involved is necessary in order to reproduce
 correctly the observed experimental data. \\
  

PACS: 0.3.65.Bz

\end{abstract}


\section{Introduction.}

One of the most characteristic features of Quantum Mechanics, the one that
 according to Schr\"odinger words {\em ``enforces its entire departure from
 classical line of thoughts''}~\cite{ref1}, is represented by the
 Entanglement.
This property, also commonly known as Quantum Non-Separability, consists in
 the impossibility of attributing to the constituent subsystems of some
 composite quantum system well-defined state vectors.
Thus, in this very frequent physical situation, which always arises as a
 consequence of an interaction between the subsystems, it is impossibile to
 ascribe them objective properties (or elements of reality).
Entangled states played an important role in the development of the Foundations
 of Quantum Mechanics since the very first times of its history: in fact
 the incompleteness argument of EPR is essentially based on the properties
 of an entangled couple of elementary particles.
Moreover, Bell's inequalities,
 proving the unavoidable non-local features of every conceivable theory which
 gives the same probabilistic predictions of the Orthodox Quantum Mechanics,
 exploit the correlations between the outcomes of measurement processes
 performed on entangled states.
In the present days the interest of the scientific community towards the
 entanglement has considerably increased since it has been acknowleged as
 an essential resource in the development of the quantum theories of
 Information and Computation.
In fact, the possibility of performing successful teleportation procedures
 of arbitrary
 and unknown quantum states~\cite{ref2}, the possibility of transmitting two
 bits of classical information by manipulating only one qubit~\cite{ref3},
 and the possibility of implementing quantum algorithms able to solve in a more
 efficient way hard computational problems~\cite{ref4}, are mainly due
 to the non-local correlation properties displayed by the entangled states and
 constitute only a few of a larger number of examples which undoubtly
 prove the usefulness of entanglement.
In this paper we focus our attention on the analysis of the mathematical
 description of the first experimental realization of the quantum
 teleportation process~\cite{ref5,ref6}, in which an arbitrary
 polarization state carried by a photon is transmitted and later reconstructed
 over a distant spatial location by means of an entangled quantum channel,
 corresponding to two more other photon particles.
In order to reproduce correctly the observed experimental data, we will see tha
t 
 a symmetrization procedure over both the polarization and spatial
 degrees of freedom of all the indistinguishable particles involved is needed.
The analysis of the experiment shows that the
 identity of the particles, and the consequent and compulsory introduction of
 the spatial part of the Hilbert space, which is not present in the original
 work of Bennett et al.~\cite{ref2} since no explicit mention was made about
 the nature of the particles, has not been exhaustively stressed in the usual
 discussions about the subject. 


\section{Usual teleportation scheme.}

The purpose of the teleportation process consists in making a perfect copy of
 an arbitrary and even unknown quantum state at a distant location, in
 accordance with the special relativity principles and the no-cloning theorem.
This process is succesfully achieved by means of the peculiar features
 displayed by certain maximally entangled quantum states, which exhibit
 non-local correlations between the outcome results of measurement operations.
The experiment described in~\cite{ref5} succeeded in
 realizing a teleportation scheme permitting the transmission of an arbitrary
 polarization state of a photon between two distant spatial regions.
It involvs the production of a couple of entangled photons by means
 of a parametric down conversion process, a particular kind of Bell-measurement
 performed by a beamsplitter used as a Bell-state analyzer and, finally, a
 classical
 communication between the two distant parties, which allows the receiver to
 successfully reconstruct the initial quantum state.

Before analyzing this experiment, it is worthwhile reviewing briefly the 
 main mathematical ingredients of the original teleportation scheme devised by
 Bennett et al.~\cite{ref2}.
First of all we note that they make the tacit assumption that the quantum
 physical systems involved are not identical, the global state vector
 representing the whole system being not symmetrized or antisymmetrized.

In fact, they consider an unknown two-level quantum state $\vert \psi
 \rangle_{1}= \alpha \vert 0 \rangle_{1} +\beta \vert 1\rangle_{1}$ ($\,$where
 $\left\{ \vert 0 \rangle,\vert 1\rangle \right\}$ are a
 complete set of orthonormal vectors$\,$) coupled with a maximally entangled
 singlet state of two other quantum systems constituting a
 so-called quantum channel, $\vert \omega \rangle_{23}=1/ \sqrt{2}\, [\,
 \vert 01\rangle_{23}-\vert 10\rangle_{23}\,]$.
By performing a Bell-state measurement on the initial state $\vert \psi
 \rangle_{1} \vert \omega \rangle_{23}$ onto the four maximally
 entangled vectors describing the first and the second particle
\begin{equation}
\label{basedibell}
\left\{
\begin{array}{lll}
\vert \phi^{\pm} \rangle_{12} & = & \frac{1}{\sqrt{2}}\,[\,\vert 00\rangle_{12}
 \pm \vert 11\rangle_{12}\,], \\
\vert \psi^{\pm} \rangle_{12} & = & \frac{1}{\sqrt{2}}\,[\,\vert 01\rangle_{12}
 
\pm \vert 10\rangle_{12}\,],
\end{array}
\right.
\end{equation}
one obtains four possible and equiprobable wave function reduction processes,
 whose resulting quantum states are:
\begin{equation}
\begin{array}{lll}
\vert \psi \rangle_{1} \vert \omega \rangle_{23} & \rightarrow \:\:&
\vert \phi^{+} \rangle_{12}\,[\,-\beta \vert 0 \rangle_{3} +
 \alpha \vert 1\rangle_{3}\,],\\
 & \rightarrow \:\:&
\vert \phi^{-} \rangle_{12}\,[\,+\beta \vert 0 \rangle_{3} +
 \alpha \vert 1\rangle_{3}\,],\\
& \rightarrow \:\:&
\vert \psi^{+} \rangle_{12}\,[\,-\alpha \vert 0 \rangle_{3} + 
 \beta \vert 1\rangle_{3}\,],\\
& \rightarrow \:\:&
\vert \psi^{-} \rangle_{12}\,[\,- \alpha \vert 0 \rangle_{3} -
 \beta \vert 1\rangle_{3}\,].
\end{array}
\end{equation}
Knowing the Bell-state which has been obtained,
one immediately sees that it is possible to get the initial quantum state
 $\vert \psi \rangle$, which had to be teleported, by performing 
 suitable unitary manipulations on the third particle.
 
However, this mathematical treatment of the teleportation process ceases to be
 valid when considering quantum states which describe indistinguishable
 physical systems, this situation being the one described in~\cite{ref5}.
In fact, in such a case the quantum mechanical formalism for identical
 particles forces one to consider state vectors with definite symmetry
 properties under arbitrary permutations of the three particles involved.
Although this mathematical requirement appears as obvious, no care is given
 to the experimental impossibility of distinguishing between particle one,
 two and three.
This indistinguishability descends from the fact that, both in the
 parametric down conversion process and in the Bell-state measurement within
 the analyzer, the wave functions of all the particles overlap each other.
Moreover, as clearly stated in~\cite{ref5}, the considered experimental set up
 is able to perform a Bell-measurement onto the only Bell state
 $\vert \psi^{-} \rangle$, which is discriminated from the other
 three by means of a position measurement within an interferometric apparatus.
It is therefore necessary to introduce explicitly also the spatial part of the
 state vector describing the whole system of the three photon particles, in
 addition to its polarization part, and the symmetrization procedure must be
 extended to these degrees of freedom too.

It is our purpose to show that, if one follows the description of the
 teleportation experiment given in~\cite{ref5}, it is not possible to perform
 a successful reconstruction of an unknown polarization state of a photon, this
 
 fact being due to the assumption of the distinguishability of
 one of the three particles involved and to the absence of the spatial part of
 the state vectors of all the particles, part which should be present from the
 very beginning of the mathematical description.In fact, the initial vector 
 describing the three photons is once again
 the one chosen by Bennett et al., i.e. the one
 composed of the polarization state $\vert \psi \rangle_{1}$ of particle 1,
 which must be teleported, and of an entangled couple of photons described by
 the usual singlet state $\vert \omega \rangle_{23}$.
It can be written in a more suitable way in terms of the vectors belonging to
 the Bell basis of equations (2.1) as follows:
\begin{eqnarray}
\vert \psi \rangle_{1} \vert \omega \rangle_{23}& = & \frac{1}{2}\,\Big[\,
\vert \phi^{+} \rangle_{12}\,[\,-\beta \vert 0 \rangle_{3} +
 \alpha \vert 1\rangle_{3}\,] +
\vert \phi^{-} \rangle_{12}\,[\,\beta \vert 0 \rangle_{3} +
 \alpha \vert 1\rangle_{3}\,]\nonumber \\
&&\:\:\:\:\:\:\:
\vert \psi^{+} \rangle_{12}\,[\,-\alpha \vert 0 \rangle_{3} + 
 \beta \vert 1\rangle_{3}\,] -
\vert \psi^{-} \rangle_{12}\,[\,\alpha \vert 0 \rangle_{3} +
 \beta \vert 1\rangle_{3}\,]\:\Big].
\end{eqnarray}    
In order to perform the Bell-measurement and complete in such a way the
 teleportation procedure, we must project this state onto one of the four
 Bell-states. This task is accomplished by means of the use of a
 beamsplitter which is able to discriminate the only state $\vert \psi^{-}
 \rangle$ from the remaining three, and this happens with
 probability 1/4.
This Bell-state analyzer acts on the spatial degrees of freedom of
 the particles 1 and 2 impinging on it, leaving unchanged their polarization
 state, and producing the following effect: if the particles are described by a
n
 antisymmetric spatial state vector, they emerge separately in two different
 outputs beyond the beamsplitter, while, if the spatial part is symmetric,
 they emerge both in the same output.     

It is only at this point of the treatment that the authors of~\cite{ref5}
 consider that the
 two photons 1 and 2 impinging onto the beamsplitter are identical particles
 and, consequently, that the Bell vectors must display an overall symmetry
 property on the polarization and the spatial degrees of freedom:
\begin{eqnarray}
\label{baseampliata}
\vert \psi^{+}_{\cal{S}} \rangle & = & \frac{1}{2}
[\, \vert 0\rangle \vert 1\rangle +\vert 1\rangle\vert 0\rangle\,]
[\, \vert A\rangle \vert B\rangle +\vert B\rangle\vert A\rangle\,],\nonumber \\
\vert \psi^{-}_{\cal{S}} \rangle & = & \frac{1}{2}
[\, \vert 0\rangle \vert 1\rangle -\vert 1\rangle\vert 0\rangle\,]
[\, \vert A\rangle \vert B\rangle -\vert B\rangle\vert A\rangle\,],\nonumber \\
\vert \phi^{+}_{\cal{S}} \rangle & = & \frac{1}{2}
[\, \vert 0\rangle \vert 0\rangle +\vert 1\rangle\vert 1\rangle\,]
[\, \vert A\rangle \vert B\rangle +\vert B\rangle\vert A\rangle\,],\nonumber \\
\vert \phi^{-}_{\cal{S}} \rangle & = & \frac{1}{2}
[\, \vert 0\rangle \vert 0\rangle -\vert 1\rangle\vert 1\rangle\,]
[\, \vert A\rangle \vert B\rangle +\vert B\rangle\vert A\rangle\,],\nonumber \\
\end{eqnarray}
where we have indicated with $\vert A\rangle$ and $\vert B\rangle$ two non
 overlapping (i.e. orthogonal) spatial regions. A closer inspection of these
 states reveals that a joint detection of particles 1 and 2 by two detectors
 placed in two different outputs beyond the beamsplitter, identify uniquely,
 according to what we have said before, the second of the above states, being
 the only one described by an antisymmetric spatial part.
The authors therefore conclude that only in this case it is possible to 
 affirm that the teleportation procedure has been successfully achieved,
 and that the polarization part of photon 3, after the measurement, is
 described by the state $\alpha \vert 0\rangle + \beta \vert 1\rangle$.

We claim that this procedure is clearly not correct, since photon 2 is 
 considered distinguishable from photon 3, even if they have been
 produced in a parametric down
 converter which undoubtly has mixed their spatial wave functions.

Let us try to reformulate the whole analysis by considering particles 2
 and 3 indistinguishable and by adding the spatial
 degrees of freedom for all the particles.   
We then consider the following normalized state vector
 $\vert \psi \rangle_{1}$,
 describing a photon located in the bounded spatial region
 $\vert A \rangle$, whose polarization state we want to teleport to another
 distant spatial region $\vert C \rangle$:
\begin{equation}
\label{tobeteleported}
\vert \psi \rangle_{1}= [\, \alpha\vert 0\rangle_{1} +\beta\vert 1
\rangle_{1}\,]\,\vert A \rangle_{1}\:\:,
\end{equation}
 and the quantum channel $\vert \phi \rangle$ consisting again of an entangled 
 couple of indistinguishable photons, whose describing vector however
 displays explicitly the
 overall symmetry property both on the polarization and on the spatial degrees
 of freedom:
\begin{equation}
\label{channeltwo}
\vert \phi \rangle_{23}= \frac{1}{2}\,[\,\vert 0 \rangle_{2}\vert 1 \rangle_{3}
- \vert 1 \rangle_{2}\vert 0 \rangle_{3}\,]
[\,\,\vert B \rangle_{2}\vert C \rangle_{3}- \vert C \rangle_{2}\vert B 
\rangle_{3}\,]\:.
\end{equation}
The polarization part of the state $\vert \phi \rangle_{23}$ is the usual
 singlet state, while the spatial part describes two particles having reached
 two distant and non-overlapping regions $\vert B\rangle$ and $\vert C\rangle$.
This state is clearly totally entangled, since we cannot attribute any
 objective property to the single photons: we can only say that the
 probability of finding a particle
 ($\,$without being able to say which one of the two$\,$) with a
 particular polarization ($\,$$\vert 0 \rangle$ or $\vert 1\rangle$$\,$) in a
 precise region of space ($\,$$\vert B \rangle$ or $\vert C\rangle$$\,$) is
 always $1/2$.

The initial state vector of the three particles is therefore the following one:
\begin{eqnarray}
\label{statoiniziale}
\vert \Omega \rangle_{123}=\vert \psi \rangle_{1}\vert \phi \rangle_{23}& =
\frac{1}{2}& \Big[
 \alpha \vert 0A\rangle_{1}\vert 0B\rangle_{2}\vert 1C\rangle_{3}
-\alpha \vert 0A\rangle_{1}\vert 0C\rangle_{2}\vert 1B\rangle_{3}\nonumber  \\
&&+\beta \vert 1A\rangle_{1}\vert 0B\rangle_{2}\vert 1C\rangle_{3}
-\beta \vert 1A\rangle_{1}\vert 0C\rangle_{2}\vert 1B\rangle_{3} \nonumber \\
&&-\alpha \vert 0A\rangle_{1}\vert 1B\rangle_{2}\vert 0C\rangle_{3}
+\alpha \vert 0A\rangle_{1}\vert 1C\rangle_{2}\vert 0B\rangle_{3}\nonumber \\
&&-\beta \vert 1A\rangle_{1}\vert 1B\rangle_{2}\vert 0C\rangle_{3}
+\beta \vert 1A\rangle_{1}\vert 1C\rangle_{2}\vert 0B\rangle_{3} \Big].
\end{eqnarray}
Incidentally we note that in the state (\ref{statoiniziale}) the photon
 labelled with the index 1 is kept distinguishable from the remaining two
 until it impinges upon the beamsplitter.

In order to complete the teleportation procedure by a measurement operation,
 it is again necessary to express the states of the photons located in regions
 $A$ and $B$ in terms of the Bell-states~(\ref{baseampliata}).
Yet, these states are no longer sufficient to express
 completely the state (\ref{statoiniziale}), since it has no symmetry
 properties with respect to the exchange of particles 1 and 2. 
In order to overcome this
 difficulty, one has to resort to other four states:
\begin{eqnarray}
\label{baseampliata2}
\vert \psi^{+}_{\cal{A}} \rangle & = & \frac{1}{2}
[\, \vert 0\rangle \vert 1\rangle +\vert 1\rangle\vert 0\rangle\,]
[\, \vert A\rangle \vert B\rangle -\vert B\rangle\vert A\rangle\,],\nonumber \\
\vert \psi^{-}_{\cal{A}} \rangle & = & \frac{1}{2}
[\, \vert 0\rangle \vert 1\rangle -\vert 1\rangle\vert 0\rangle\,]
[\, \vert A\rangle \vert B\rangle +\vert B\rangle\vert A\rangle\,],\nonumber \\
\vert \phi^{+}_{\cal{A}} \rangle & = & \frac{1}{2}
[\, \vert 0\rangle \vert 0\rangle +\vert 1\rangle\vert 1\rangle\,]
[\, \vert A\rangle \vert B\rangle -\vert B\rangle\vert A\rangle\,],\nonumber \\
\vert \phi^{-}_{\cal{A}} \rangle & = & \frac{1}{2}
[\, \vert 0\rangle \vert 0\rangle -\vert 1\rangle\vert 1\rangle\,]
[\, \vert A\rangle \vert B\rangle -\vert B\rangle\vert A\rangle\,],\nonumber \\
\end{eqnarray}
However, these new states are totally antisymmetric and therefore cannot
 correspond to the physical states of the photons which enter the
 interferometric apparatus of the Bell-state analyzer. 

In any case, the two sets of vectors (\ref{baseampliata}) and
 (\ref{baseampliata2}) taken together are necessary and sufficient in order
 to express the state of the particles located in regions $\vert A \rangle$ and
 $\vert B \rangle$ of equation~(\ref{statoiniziale}) in terms of vectors with
 a spatial part with definite symmetry. In fact we have:
\begin{eqnarray}
\vert 0A \rangle \vert 0B\rangle &=& \frac{1}{2}\,[\,
\vert \phi^{+}_{\cal{S}} \rangle +\vert \phi^{-}_{\cal{S}} \rangle
+\vert \phi^{+}_{\cal{A}} \rangle +\vert \phi^{-}_{\cal{A}} \rangle\,]\:,
\nonumber \\
\vert 0A \rangle \vert 1B\rangle &=& \frac{1}{2}\,[\,
\vert \psi^{+}_{\cal{S}} \rangle +\vert \psi^{-}_{\cal{S}} \rangle
+\vert \psi^{+}_{\cal{A}} \rangle +\vert \psi^{-}_{\cal{A}} \rangle\,]\:,
\nonumber \\
\vert 1A \rangle \vert 0B\rangle &=& \frac{1}{2}\,[\,
\vert \psi^{+}_{\cal{S}} \rangle -\vert \psi^{-}_{\cal{S}} \rangle
+\vert \psi^{+}_{\cal{A}} \rangle -\vert \psi^{-}_{\cal{A}} \rangle\,]\:,
\nonumber \\
\vert 1A \rangle \vert 1B\rangle &=& \frac{1}{2}\,[\,
\vert \phi^{+}_{\cal{S}} \rangle -\vert \phi^{-}_{\cal{S}} \rangle
+\vert \phi^{+}_{\cal{A}} \rangle -\vert \phi^{-}_{\cal{A}} \rangle\,]\:.
\end{eqnarray}
If we insert these vectors into equation (\ref{statoiniziale}), a
 straightforward rearrangement gives:
\begin{eqnarray}
\label{riarrangiato}
\vert \Omega \rangle_{123} & = & \frac{1}{4} \Big[\:
[\,\vert \phi^{+}_{\cal{S}} \rangle_{12} +
 \vert \phi^{+}_{\cal{A}} \rangle_{12}\,]
\,[\,\alpha \vert 1C\rangle_{3} -\beta \vert 0C\rangle_{3}\,] +
[\,\vert \phi^{-}_{\cal{S}} \rangle_{12} +
 \vert \phi^{-}_{\cal{A}} \rangle_{12}\,]
\,[\,\alpha \vert 1C\rangle_{3} +\beta \vert 0C\rangle_{3}\,] \nonumber \\
&& \,
\:\:+[\,\vert \phi^{+}_{\cal{S}} \rangle_{13} +
 \vert \phi^{+}_{\cal{A}} \rangle_{13}\,]
\,[\,\alpha \vert 1C\rangle_{2} -\beta \vert 0C\rangle_{2}\,] +
[\,\vert \phi^{-}_{\cal{S}} \rangle_{13} +
 \vert \phi^{-}_{\cal{A}} \rangle_{13}\,]
\,[\,\alpha \vert 1C\rangle_{2} +\beta \vert 0C\rangle_{2}\,] \nonumber \\
&&\,
\:\:-[\,\vert \psi^{+}_{\cal{S}} \rangle_{12} +
 \vert \psi^{+}_{\cal{A}} \rangle_{12}\,]\,
[\,\alpha \vert 0C\rangle_{3} -\beta \vert 1C\rangle_{3}\,] -
[\,\vert \psi^{-}_{\cal{S}} \rangle_{12} +
 \vert \psi^{-}_{\cal{A}} \rangle_{12}\,]
\,[\,\alpha \vert 0C\rangle_{3} +\beta \vert 1C\rangle_{3}\,] \nonumber \\
&&\,
\:\:-[\,\vert \psi^{+}_{\cal{S}} \rangle_{13} +
 \vert \psi^{+}_{\cal{A}} \rangle_{13}\,]\,
[\,\alpha \vert 0C\rangle_{2} -\beta \vert 1C\rangle_{2}\,] -
[\,\vert \psi^{-}_{\cal{S}} \rangle_{13} +
 \vert \psi^{-}_{\cal{A}} \rangle_{13}\,]
\,[\,\alpha \vert 0C\rangle_{2} +\beta \vert 1C\rangle_{2}\,]\:\:. \nonumber \\
&&
\end{eqnarray}
Omitting for a moment the fact that part of the Bell-states are non physical,
 it is instructive to see why it is not possible to perform a
 successful teleportation in region $C$ of the polarization state
 $\alpha \vert 0 \rangle +\beta \vert 1\rangle$.
A joint detection of particle 1 and of one of the remaining particles
 ($\,$2 or 3, being they indistinguishable$\,$) in the two different outputs
 located beyond the beamsplitter, causes the collapse onto the state:
\begin{eqnarray}
\vert \Omega \rangle_{123} & \rightarrow \frac{1}{2\sqrt{2}} &
 \Big[\,
 -\vert \psi^{-}_{\cal{S}} \rangle_{12}\,[\,\alpha \vert 0C\rangle_{3}
 +\beta \vert 1C\rangle_{3}\,] -
\vert \psi^{-}_{\cal{S}} \rangle_{13}\,[\,\alpha \vert 0C\rangle_{2}
 +\beta \vert 1C\rangle_{2}\,] \nonumber \\
&& \:\:\:\:+\,
\vert \phi^{+}_{\cal{A}}\rangle_{12}\,[\,\alpha \vert 1C\rangle_{3} -\beta\vert
0C\rangle_{3}\,] +
\vert \phi^{+}_{\cal{A}}\rangle_{13}\,[\,\alpha \vert 1C\rangle_{2} -\beta\vert
0C\rangle_{2}\,] \nonumber \\
&& \:\:\:\: +\,
\vert \phi^{-}_{\cal{A}}\rangle_{12}\,[\,\alpha \vert 1C\rangle_{3} +\beta\vert
0C\rangle_{3}\,] +
\vert \phi^{-}_{\cal{A}}\rangle_{13}\,[\,\alpha \vert 1C\rangle_{2} +\beta\vert
0C\rangle_{2}\,]  \nonumber \\
&&\:\:\:\:
-\vert \psi^{+}_{\cal{A}}\rangle_{12}\,[\,\alpha \vert 0C\rangle_{3} 
-\beta\vert 1C\rangle_{3}\,] 
-\vert \psi^{+}_{\cal{A}}\rangle_{13}\,[\,\alpha \vert 0C\rangle_{2}
 -\beta\vert 1C\rangle_{2}\,]\: \Big],
\end{eqnarray}
where all the kets describing particles in $A$ and $B$ have an
 antisymmetric spatial part.
It is therefore apparent that, also in this non physical situation, we cannot
 conclude anything about the polarization part of the photon located in region
 $\vert C\rangle$, and that the teleportation process cannot be achieved.

We conclude this section by stressing that it is not correct to forget the
 indistinguishability of particles 2 and 3. However, even if we take into
 account that they are identical
 quantum systems, and we add the spatial degrees of freedom from the 
 beginning, it is not possible to express the global state only in terms of 
 the symmetric Bell-states~(\ref{baseampliata}).


\section{Teleportation with indistinguishable particles.}

We show now that a correct symmetrization procedure performed over both the
 spatial and the polarization degrees of freedom of all the particles involved,
 allows one to achieve a successful teleportation process, at least in $1/4$ of
 all the cases, in complete agreement with the tested experimental
 data.
Therefore, we assume as initial state vector of the teleportation scheme
 the state

\begin{equation}
\vert \tilde{\Omega} \rangle_{123} =
 {\cal{S}}\,\Big[\: \vert \psi \rangle_{1}\vert \phi \rangle_{23}\:\Big]\:,
\end{equation}

where the operator ${\cal{S}}$ performs a complete symmetrization over all the
 particles and makes the state totally symmetric.
In such a way the symmetric vectors~({\ref{baseampliata}) become again
 sufficient for describing the states of
 the particles located in regions $\vert A\rangle$ and $\vert B\rangle$.
Inserting the set of vectors~({\ref{baseampliata}) into expression (3.1),
 the state vector of the system exhibits the following mathematical form:

\begin{eqnarray}
\vert \tilde{\Omega} \rangle_{123}\!\!\! & =\! \frac{1}{\sqrt{12}}\!\! & 
\!\!\Big[\,
\vert \phi^{+}_{\cal{S}} \rangle_{12}
[\alpha \vert 1C\rangle_{3}-\beta \vert 0C \rangle_{3}]
+\vert \phi^{+}_{\cal{S}} \rangle_{13}
[\alpha \vert 1C\rangle_{2}-\beta \vert 0C \rangle_{2}]
+\vert \phi^{+}_{\cal{S}} \rangle_{23}
[\alpha \vert 1C\rangle_{1}-\beta \vert 0C \rangle_{1}] \nonumber \\
& & \!\!\!
+\vert \phi^{-}_{\cal{S}} \rangle_{12}
[\alpha \vert 1C\rangle_{3}+\beta \vert 0C \rangle_{3}]
+\vert \phi^{+}_{\cal{S}} \rangle_{13}
[\alpha \vert 1C\rangle_{2}+\beta \vert 0C \rangle_{2}]
+\vert \phi^{+}_{\cal{S}} \rangle_{23}
[\alpha \vert 1C\rangle_{1}+\beta \vert 0C \rangle_{1}] \nonumber \\
& &\!\!\!
+\vert \psi^{+}_{\cal{S}} \rangle_{12}
[\beta \vert 1C \rangle_{3}-\alpha \vert 0C\rangle_{3}]
+\vert \psi^{+}_{\cal{S}} \rangle_{13}
[\beta \vert 1C \rangle_{2}-\alpha \vert 0C\rangle_{2}]
+\vert \psi^{+}_{\cal{S}} \rangle_{23}
[\beta \vert 1C \rangle_{1}-\alpha \vert 0C\rangle_{1}] \nonumber \\
& &\!\!\!
-\vert \psi^{-}_{\cal{S}} \rangle_{12}
[\alpha \vert 0C\rangle_{3} +\beta \vert 1C \rangle_{3}]
-\vert \psi^{-}_{\cal{S}} \rangle_{13}
[\alpha \vert 0C\rangle_{2} +\beta \vert 1C \rangle_{2}]
-\vert \psi^{-}_{\cal{S}} \rangle_{23}
[\alpha \vert 0C\rangle_{1} +\beta \vert 1C \rangle_{1}]\,\Big]\:.\nonumber \\
& &
\end{eqnarray}

Owing to the same considerations we have done in dealing with the previous
 teleportation scheme, we can state that a joint detection of two
 particles in the two different outputs beyond the beamsplitter
 ($\,$this case occurring exactly in $1/4$ of the cases$\,$)
 forces the state to collapse onto the following manifold:

\begin{equation}
\vert \tilde{\Omega} \rangle_{123} \rightarrow  \frac{1}{\sqrt{3}} \Big[\,
\vert \psi^{-}_{\cal{S}} \rangle_{12}
(\alpha \vert 0C\rangle_{3} +\beta \vert 1C \rangle_{3})+
\vert \psi^{-}_{\cal{S}} \rangle_{13}
(\alpha \vert 0C\rangle_{2} +\beta \vert 1C \rangle_{2})+
\vert \psi^{-}_{\cal{S}} \rangle_{23}
(\alpha \vert 0C\rangle_{1} +\beta \vert 1C \rangle_{1})\, \Big].
\end{equation}

The polarization state describing the particle located in region
 $\vert C\rangle$ (without knowing which of the three it really is, due
 to their indistinguishability) has exactly the same form as the one we want
 to teleport.
We can therefore state that {\em in region $C$ there is with certainty a photon
 whose polarization state is $\alpha \vert 0 \rangle +\beta \vert 1 \rangle$}. 
On the contrary of what has happened in the previous section the
 teleportation scheme has been successfully realized thanks to the correct
 symmetrized form of the the state describing the three particles, which
 cannot by no means be considered distinguishable.


\section{Conclusions.}

With the aim of sheding light on the mathematical treatment
 of the teleportation experiment involving identical particles, like the one
 described in~\cite{ref5}, we have shown that a correct formulation of the
 subject, which turns out to be in perfect agreement with the observed
 experimental data, must include a symmetrization procedure over all the
 degrees of freedom of the particles involved.
In fact, in the case of a teleportation procedure of an unknown polarization
 state
 of a photon, both the parametric down conversion process and the subsequent
 Bell-state analysis within an interferometer, forces the three particles to
 become definitely indistinguishable, due to the unavoidable overlapping of
 their spatial wavefunctions during the interaction.
Moreover, in the described scheme the Bell-measurement is performed through a
 joint position detection of two particles beyond a beamsplitter.
These two considerations necessarily imply that the spatial part of the 
 wavefunction of the particles must be
 included explicitly in the initial state, and that the symmetrization
 procedure must be performed over both the polarizarition and the spatial
 degrees of freedom.
Only in this way the theoretical treatment of the teleportation process is
 able to predict correctly the observed experimental behaviour.



\end{document}